\documentstyle[12pt]{article}
\begin{document}

\hsize=16.5truecm
\addtolength{\topmargin}{-0.3in}
\addtolength{\textheight}{1in}
\vsize=26truecm
\hoffset=-.5in

\thispagestyle{empty}
\begin{flushright} July \ 1996\\
MIT-CTP-2553/SNUTP 96-080\\
\end{flushright}
\begin{center}
 {\large\bf  Scattering of Solitons  \\ in Derivative Nonlinear Schr\"{o}dinger Model  }
\vglue .5in
Hyunsoo Min
\vglue .2in
{\it
 Department of Physics \\
Seoul City University \\
Seoul, 130-743, Korea }
\vglue .2in
{and}
\vglue .2in
 Q-Han Park\footnote{ E-mail address; qpark@nms.kyunghee.ac.kr }
\vglue .2in
{\it  
Department of Physics \\
and \\
Research Institute of Basic Sciences \\
Kyunghee University\\
Seoul, 130-701, Korea}
\vglue .6in
{\bf ABSTRACT}\\[.2in]
\end{center}
We show that the chiral soliton model recently introduced by Aglietti et al. can be 
made integrable by adding an attractive potential with a fixed coefficient. 
The modified model is equivalent to the derivative nonlinear Schr\"{o}dinger model 
which does not possess parity and Galilean invariance.  
We obtain explicit one and two classical soliton solutions and show that in the 
weak coupling limit, they correctly reproduce the bound state energy as well as  
the time delay of two-body quantum mechanics of the model.

\vglue .1in

\newpage
It has been known for sometime that soliton solutions to certain nonlinear 
field equations can be associated with  elementary particles in quantum field theory. In particular, the  nonlinear Schr\"{o}dinger system shows that 
its classical solitons behave as  quantized particles of the same 
theory. This may be compared with the sine-Gordon solitons which are 
associated with quantized elementary particles of  the massive Thirring model.
Recently, this correspondence has been extended to a new type of 1+1-dim 
nonlinear field equation\cite{Jackiw,Rabello}, which  arose in the study of the dimensionally reduced 2+1-dim nonlinear Schr\"{o}dinger model coupled to Chern-Simons gauge theory\cite{Pi},  breaking the Galilean invariance.  
It has been noted that this theory 
supports a chiral soliton solution, whose mass formula  justifies the 
interpretation of a soliton as a bound state of elementary particles of the 
quantized theory in the weak coupling limit\cite{Jackiw}.  
However, one serious drawback 
of the theory was its lacking of integrability structure which made 
impossible of finding multi-soliton solutions and  their subsequent  
scattering behaviors. 
In this Letter, we show that this problem can be cured nicely when 
we add an attractive potential term to the theory with a fixed coefficient. 
This allows us to find exact one and two soliton solutions and address their 
quantum mechnical particle-like behavior. We find that one soliton 
solution reproduces the bound state energy of two identical particles,  
and reduces to the one soliton solution of Ref.\cite{Jackiw} in the weak 
coupling limit. Two soliton solution describes the soliton-soliton scattering 
from which we obtain the time delay of each solitons during the scattering 
process. It is shown that these time delays agree with those of quantum 
particle scattering thereby confirming the quantum particle interpretation 
of solitons of the present model.

The model we consider is given by the Lagrangian,
\begin{equation}
L=\int dx \Bigg[i\hbar \Psi^* \partial_t \Psi - {\hbar^2 \over 2 m } 
| (\partial_x \mp i\kappa^2 \rho)\Psi|^2 + b {\hbar^2\kappa^4 \over 2m} 
\rho^3 \Bigg]
\label{lagr}
\end{equation}
where $\rho=|\Psi|^2$,  $\kappa^2 $  is the coupling constant  and $b$ is a 
number to be fixed later.  
Note that this Lagrangian is not invariant under parity but  becomes 
invariant if we combine parity with changing the sign of the coupling 
constant,
\begin{equation}
x \rightarrow -x, \ \ \ \kappa^2 \rightarrow  - \kappa^2 . 
\end{equation}
Thus from now on, we will restrict only to the upper sign case.  
The lower sign case can be obtained directly by changing $x$ to $-x$ in 
the upper sign case.
The equation of motion  is 
\begin{eqnarray}
i\hbar  \partial_t \Psi &=& - {\hbar^2 \over 2 m } (\partial_x -i\kappa^2 
\rho)^2\Psi - \hbar\kappa^2 j \Psi
- 3b {\hbar^2\kappa^4 \over 2m} \rho^2 \Psi  \nonumber \\
&=&- {\hbar^2 \over 2 m } \partial_x^2\Psi +2i{\hbar^2 \kappa^2\over m } 
\rho \partial_x \Psi - {3\over 2}(b-1) { \hbar^2 \kappa^4\over m} \rho^2 
\Psi .
\label{eqmot}
\end{eqnarray}
The current $j$,  defined by
\begin{equation} 
j(x)={\hbar \over  2i m}\Big[ \Psi^*(\partial_x-i \kappa^2\rho )\Psi 
- \Psi (\partial_x + i \kappa^2 \rho )\Psi^*  ) \Big] ,
\label{current}
\end{equation}
 satisfies the continuity equation
\begin{equation}
\partial_{t}\rho + \partial_{x} j =0 
\label{cont}
\end{equation}
when the equation of motion (\ref{eqmot}) is used.  
The $b=0$ case, i.e. without the 
potential term in the Lagrangian (\ref{lagr}), was considered in Ref.\cite{Jackiw} 
and one soliton solution with chiral behavior has been obtained. 
However, the Painlev\'{e} test shows that  Eq.(\ref{eqmot}) becomes 
integrable only for the value $b =1$  \cite{Clarkson}, in which case the 
equation is known as the derivative nonlinear Schr\"{o}dinger equation 
of type II. Henceforth, we will set $b=1$. In order to compare with 
other literatures concerning about the derivative nonlinear 
Schr\"{o}dinger equation, we may take a transform of 
$\Psi $ with an arbitrary constant $a$, 
\begin{equation}
\Psi=e^{-i a \kappa^2\int^x  dy \rho(y)}  \psi , 
\end{equation}
which brings Eq.(\ref{eqmot}) into a different, yet equivalent expression,
\begin{equation}
i\hbar  \partial_t \psi
=- {\hbar^2 \over 2 m } \partial_x^2\psi  +  i \kappa^2{\hbar^2 \over m } 
[(2+a)\rho \partial_x \psi + a\psi^2\partial_x\psi^* ]
+ { \hbar^2\kappa^4 \over 2 m}a (2-a)\rho^2 \psi ,
\label{eqmot2}
\end{equation}
where $\rho = |\Psi |^{2} = |\psi |^2 $. 
For $a=-1$, we recover the expression of $j  \psi $-coupling given in 
Ref.\cite{Jackiw}  with an additional term, 
\begin{equation}
i\hbar  \partial_t \psi =
- {\hbar^2 \over 2 m } \partial_x^2\psi -2 \hbar \kappa^2   j \psi
- {3 \hbar^2 \kappa^4  \over 2 m}\rho^2 \psi.
\end{equation}
For $a=2$,   Eq.(\ref{eqmot2}) reduces to the derivative nonlinear 
Schr\"{o}dinger  equation of  type I;
 \begin{equation}
i\hbar  \partial_t \psi =
- {\hbar^2 \over 2 m } \partial_x^2\psi +2i{ \hbar^2 \kappa^2 \over m }
\partial_x(\rho  \psi).
\end{equation}
For the sake of brevity, we introduce dimensionless coordinates 
\begin{equation}
X=4 \kappa^2x  \ \ , \ \ \  T= 8( \hbar\kappa^4 /m ) t 
\end{equation}
and express Eq.(\ref{eqmot}) with $b=1$ in a dimensionless form,
\begin{equation}
i\partial_{T}\Psi = -\partial_{X}^2 \Psi + i\rho \partial_{X}\Psi .
\label{eqmot3}
\end{equation}
A general scheme for solving the derivative nonlinear Schr\"{o}dinger 
equation has been given by Kaup and Newell in terms of the inverse 
scattering method\cite{Kaup}. Following 
their method, we construct explicitly one and 
two soliton solutions  of Eq.(\ref{eqmot3}) in the following. 
The one soliton solution is given by
\begin{equation}
\Psi = \sqrt{\rho}  \exp (- i \Theta+{i \over 4} \int^X \rho),
\label{onesol}
\end{equation}
where 
\begin{eqnarray}
\rho&=&{ 2 \mu^2 |V| \over \sqrt{1 +\mu^2 }{\rm  cosh} ( \mu V[ X - VT] ) 
+ {\rm Sign} (V)}   \nonumber  \\
\Theta&=&-{1\over 2}V[X-VT] - {1\over 4} ( 1 + \mu^2)  V^2 T + \theta .
\label{Theta}
\end{eqnarray}
$V$ is the velocity of the soliton and $\theta $ is an arbitrary initial 
phase. The parameter $\mu \ge 0 $ accounts for the mass of the soliton as 
shown below. For finite $\mu $, note that the soliton solution $\Psi $ 
vanishes when the velocity $V$ is set to zero. This means that there is 
no static finite mass soliton (with time periodic phase) thus one can 
not simply obtain the moving one soliton solution by boosting the    
static one as in the case of  the usual nonlinear Schr\"{o}dinger equation.
This clearly shows that the system is lacking of Galilean invariance.
Nevertheless, it is intriguing to see that, for infinite $\mu $ such that 
$\mu V= \alpha $ is finite, Eq.(\ref{onesol}) admits a static soliton,
\begin{equation}
\Psi e^{ - { i \over 4} \int^X  \rho}= \Big[ { 2\alpha \over \cosh(\alpha X) } \Big]^{1 \over 2} 
\exp(i\alpha^2 T/4 ) .
\end{equation}  
 For later use, we present an equivalent expression of Eq.({\ref{onesol}) 
which is useful in comparison with the two soliton scattering process,
\begin{equation}
\Psi e^{ - { i \over 2} \int^X  \rho}={\mu |V|  \over \sqrt {|\lambda |}}
{\exp(\Delta ) \exp(-i \Theta )  \over { \exp(2 \Delta ) - \lambda / |\lambda | }}
\label{another}
\end{equation} 
where
\begin{equation}
\Delta = {1 \over 2}\mu |V|  [X - VT],  \ \   \lambda={1\over 4 } 
(- V + i \mu |V|).
\label{Delta}
\end{equation}
The system possesses infinitely many conserved charges among which 
the first few are playing the role of dynamical parameters in the 
context of particle interpretation. In terms of dimensionful parameters, we evaluate those quantities explicitly on the one soliton solution with the following result; 
the constituent number $N$ is
\begin{equation}
N=\int_{-\infty }^{\infty} dx \rho={1\over \kappa^2}\tan^{-1} (\mu v/|v|)
; \ \ 0 < \kappa^{2}N < \pi
\end{equation}
while the momentum $P$ is
\begin{eqnarray}
P &= &\int^{\infty }_{-\infty } d x { \hbar \over 2 i } \Big[ \Psi^* \partial_x \Psi  - 
\partial_x \Psi^* \Psi \Big]   \nonumber \\
& = & ({\mu m \over  \kappa^2}) |v| \equiv  Mv .
\end{eqnarray}
Note that the momentum $P$ is always positive regardless of the 
sign of the velocity. The soliton mass $M$ may be written in terms of 
the constituent number $N$, 
\begin{equation}
M  =  {\mu m \over  \kappa^2}{ v \over |v|} 
={m \over \kappa^2} \tan(\kappa^2 N) . 
\end{equation}
The energy $E$, defined by
\begin{equation}
E=\int_{-\infty }^{\infty } dx \Bigg[{\hbar^2\over 2m}|(\partial_x -i \kappa^2 \rho ) 
\Psi |^2 - {\hbar^2 \kappa^4 \over 2 m}\rho^3 \Bigg]  ,     
\end{equation}
becomes
\begin{equation}
E= {1 \over 2} vP = {1 \over 2 M}  P^2 .
\end {equation}
Thus, it satisfies the particle dispersion relation.
For small $\kappa $, we may expand $M$ as
\begin{equation}
M=m (N + {1 \over 3} \kappa^4 N^3) + O(\kappa^6).
\label{mass}
\end{equation} 
In the weak coupling limit where $\kappa^2 N$ is kept small, 
the one soliton solution in Eq.(\ref{onesol}) agrees 
precisely with that of Ref.\cite{Jackiw}. However, one marked 
difference is that unlike the one soliton of Ref.\cite{Jackiw} 
which admits only positive velocity, our case admits both signs 
of  velocity. Indeed,  negative velocity in our context 
requires a strong coupling,  $\pi /2  < \kappa^2 N < \pi $, thus 
in comparing with particles in the weak coupling limit, we 
take only the positive velocity case. 

The quantum mechanics of the two-body system for our model is 
the same as that appeared in Ref.\cite{Jackiw} since the potential term 
added to the Lagrangian (\ref{lagr}) does not affect the two body 
system. Define the two-body wave function by
\begin{equation}
\Phi(x_1, x_2)=\langle 0 | \Psi(x_1) \Psi (x_2)  | 2 \rangle. 
\end{equation}
Because of time and space translational invariance, one can set
\begin{equation}
\Phi(x_1, x_2 ; t) = \exp ( - i E t / \hbar ) \exp ( i P(x_1 +x_2) / 
2 \hbar ) u(x_1 -x_2) ,
\end {equation}
so that the two-body equation for the relative motion becomes
\begin{equation}
( -{\hbar^{2} \over m}{d^2 \over dx^2 } + {P^{2} \over 4m} - 
{P \over m}\hbar \kappa^{2} \delta (x) )u(x) = Eu(x).
\end{equation}
This possesses a bound state provided  $P$ is posivtive. 
In which case, the total energy 
$ E $ is
\begin{equation}
E= {P^2 \over 4 m} (1 - \kappa^4) .
\label{massptl}
\end{equation}
On the other hand, if we assume the same type of one-loop modification 
to Eq.(\ref{mass}) as suggested in Ref.\cite{Jackiw,Nohl}, 
\begin{equation}
M_{\rm semiclassical} = mN + {1\over 3}m\kappa^4 (N^3 - N) ,
\end{equation}
we get for $N=2$
\begin{equation}
M_{\rm semiclassical} = 2m(1 + \kappa^4 )
\end{equation}
which is consistent  with Eq.(\ref{massptl}) for small $\kappa $.   

A straightforward calculation following the inverse scattering 
method for the derivative nonlinear Schr\"{o}dinger equation 
shows that the exact two soliton solution can be given by
\begin{eqnarray}
\Psi e^{ - {i \over 2} \int^X \rho}  &=& 
\Bigg[
2 e^{-\Delta_1 - i \Theta_1} +2 e^{-\Delta_2 - i \Theta_2} + 
 A_{12} e^{ -2 \Delta _1 -\Delta_2 -i\Theta_{2} } + 
A_{21} e^{ -2 \Delta _2 - \Delta_1 -i\Theta_{1} }\Bigg ]
 \nonumber \\
&&\times \Bigg[
1+D e^{ -2 \Delta _1 -2\Delta_2 } +( C_{11} e^{-2 \Delta_1  }
 + C_{12} e^{ -\Delta_1 -\Delta_2 +i \Theta_1 - i  \Theta_2}  + 
1 \leftrightarrow 2) 
\Bigg]^{-1}
\label{twosol}
\end{eqnarray}
where
\begin{eqnarray}
A_{ij} &=&  2 \lambda_{i} \Big( {1 \over \lambda_{i}-\lambda_{j}^{*} }
- {1 \over \lambda_{i} - \lambda_{i}^{*} }  \Big)^{2}    \ , \ \ 
C_{ij}= {\lambda_i \over (\lambda_i - \lambda_j^* )^2} 
      \nonumber \\
D &=& \lambda_{1}\lambda_{2} \Big( {1 \over |\lambda_1 - \lambda_2^* |^2}
 + {1 \over (\lambda_1-\lambda_1^*) ( \lambda_2 - \lambda_2^*)} \Big)^2  .
\end{eqnarray} 
 $\Theta_{1} (\Theta_{2}) $  and $\Delta_{1} (\Delta_{2})$ are as in  Eqs.(\ref{Theta}) 
and (\ref{Delta}) with velocity $V_{1}(V_{2})$.
Justification for the above expression may follow from the following asymptotic 
behavior; in the limit $ t \rightarrow \infty $ with $\Delta_{1}$ fixed 
so that $\Delta_{2} \rightarrow \infty$ for $V_{1} > V_{2}$, the two 
soliton solution in Eq.(\ref{twosol}) approaches to 
\begin{eqnarray}
\Psi e^{ - {i \over 2} \int^X \rho} &=& {  \mu_1\sqrt{|V_1|} \over \sqrt{|\lambda_1|}}
{e^{\Delta_{1} - \phi_{I} } e^{-i\Theta_{1}} \over
e^{2(\Delta_{1} - \phi_{I})} - \lambda_1  / |\lambda_1|} 
\nonumber
\\
\phi_{I} &\equiv & \ln (\sqrt{ 1+  \mu_1^2 }/\mu^2_1) 
\label{solbef}
\end{eqnarray}
which is precisely the 1-soliton in Eq.(\ref{another}) moving with 
velocity $V_{1}$. Also, another limit $t\rightarrow -\infty $ 
with $\Delta_{1} $ fixed and $\Delta_{2} \rightarrow -\infty $ results in the
1-soliton 
expression
\begin{eqnarray}
\Psi  e^{ - {i \over 2} \int^X \rho} &=&  {  \mu_1\sqrt{|V_1|}\over \sqrt{ | \lambda_1 |}}
{e^{\Delta_{1} - \phi_{F}} e^{-i\Theta_{1} + 2 i \phi_{P}} 
\over e^{2(\Delta_{1} - \phi_{F} )} -\lambda_1  / |\lambda_1|  } 
\nonumber
\\
\phi_{F} &\equiv & \phi_{I} + \ln ({|\lambda_1 - \lambda_2 |^2 \over
 |\lambda_1 - \lambda_{2}^{*} |^2 } )     \nonumber \\
\tan \phi_{P} &\equiv & 
{  (V_2 - V_1 )^2 + (\mu_1 V_1 + \mu_2 V_2 )^2 - 2 \mu_2 V_2  
(\mu_1 V_1 + \mu_2 V_2 ) 
\over 
2 \mu_2 V_2 (V_2 - V_1 )} .
\label{solaf}
\end{eqnarray}
One could repeat the same limiting procedure but with $\Delta_{2}$ fixed and
obtain the other soliton sector moving with velocity $V_{2}$. 
This shows clearly that Eq.(\ref{twosol}) is a two soliton solution 
describing the soliton-soliton scattering. 
The term $\phi_{I}$ measures the choice of time coordinate orgin $t=0$ , which 
can be simply removed by shifting the orgin. However, the difference $\phi_{F} 
-\phi_{I}$ is invariant under time translation and  measures the time delay. 
For example, the time delay of 
the soliton with velocity $V_{1}$ due to the scattering is given by,
\begin{equation}
[\Delta T]_{1} = {2 \over  \mu_1 V_1^2 } \ln  \Big[ 1 - { 4\mu_1  \mu_2 V_1 V_2 
\over 
(V_1 - V_2)^2 +  (\mu_1 V_1 + \mu_2 V_2)^2  }\Big] .
\end{equation}
In terms of dimensionful parameters, this becomes with $\mu_{1}=\mu_{2}$, 
\begin{eqnarray}
[\Delta t]_{1} &=& { \hbar \over 2 \kappa^{2} E_{1} } \ln \Big[ 1 - {
4\kappa^{4}(M/m)^{2}\sqrt{E_{1}E_{2}}
 \over (\sqrt{E_{1}} - \sqrt{E_{2}})^{2} + \kappa^{4}
(M/m)^{2}(\sqrt{E_{1}} + \sqrt{E_{2}})^{2} } \Big]     \nonumber \\
&=& -2\hbar \kappa^{2} (M/m)^{2}  { \sqrt{E_{2}} \over \sqrt{E_{1}}  }
{1 \over   (\sqrt{E_{1}} - \sqrt{E_{2}})^{2} } + {\cal O}(\kappa^{4})
\label{delay1}
\end{eqnarray}
where we have expressed velocities in terms of energies. 

In the two-body quantum mechanics, the phase shift occuring in the 
scattering of two identical particles, each having positive kinetic 
energy $E_{1}$ and  $E_{2}$, can be easily computed to give 
\begin{eqnarray}
2\delta (E_{1}, E_{2}) &=& 2 \tan^{-1}\Big( \kappa^{2}{ \sqrt{E_{1}} + 
\sqrt{E_{2}} \over
\sqrt{E_{1}} - \sqrt{E_{2}} } \Big) \nonumber \\
& = & 2\kappa^{2} { \sqrt{E_{1}} + \sqrt{E_{2}} \over \sqrt{E_{1}} -
\sqrt{E_{2}} } + {\cal O}(\kappa^{4}) .
\end{eqnarray}
Using the relation between the phase shift and the time delay\cite{Woo}, we find 
the time delay for the particle with energy $E_{1}$,
\begin{equation}
[\Delta t]_{1} = \hbar {\partial \over \partial E_{1} } 2 
\delta (E_1 , E_2 ) =
-2 \hbar \kappa^{2} { \sqrt{E_{2}} \over \sqrt{E_{1}}(\sqrt{E_{1}} -
\sqrt{E_{2}})^{2} } + {\cal O}(\kappa^{4}) .
\label{delay2}
\end{equation}
In the leading order, this agrees with Eq.(\ref{delay1}) when the 
soliton mass is taken to  be that of a quantum particle so that 
$N =1$, and the velocities are all taken positive values. This 
strengthens the quantum particle interpretation of classical 
solitons for the present case. In the case of the usual nonlinear 
Schr\"{o}dinger equation, same result has been obtained in \cite{Dolan}. 

In conclusion, we have shown that the potential term added to the 
theory of Ref.\cite{Jackiw} as in Eq.(\ref{lagr}) makes the theory 
integrable and the one and two soliton solutions correctly 
reproduce the characteristics of two-body quantum mechanics in the 
weak coupling limit. It was also shown that the theory possesses 
soliton solutions with negative velocity which do not have a 
direct particle interpretation. Since it requires strong coupling 
$ \pi /2  <  \kappa^2 N < \pi $, this seems to suggest a collective 
motion of a bound state of particles with large $N$ which however 
stays as an open problem.   
\vglue .2in
{\bf ACKNOWLEDGEMENT}
\vglue .2in
We thank R.Jackiw for his invaluable help and critical comments. 
We also thank  L. Griguolo
 and D. Seminara for discussion and the Center for Theoretical Physics at 
MIT for their warm hospitality during our visit. This work is supported in part  by 
 KOSEF through the  binational collaboration programs between KOSEF and 
NSF of USA  (H.M. with S.-Y.Pi and Q.P. with R. Jackiw ) and also through CTP/SNU.
This work is also supported in part by BSRI(Korean Ministry of Education) and  
D.O.E. under DE-FC02-94ER40818.  

\vglue .5in

\end{document}